# Tailoring the Thickness-Dependent Optical Properties of Conducting Nitrides and Oxides for Epsilon-Near-Zero-Enhanced Photonic Applications

Soham Saha*, Mustafa Goksu Ozlu*, Sarah N. Chowdhury, Benjamin T. Diroll,
Richard D. Schaller, Alexander Kildishev, Alexandra Boltasseva, and Vladimir M. Shalaev
*Equal contribution authors

Keywords: Titanium nitride, transparent conducting oxides, AZO, metasurfaces, tailorable absorbers, epsilon near zero, Berreman

## Abstract

The unique properties of the emerging photonic materials, conducting nitrides and oxides, especially their tailorability, large damage thresholds, and, importantly, the so-called epsilon-near-zero (ENZ) behavior, have enabled novel photonic phenomena spanning optical circuitry, tunable metasurfaces, and nonlinear optical devices. This work explores direct control of the optical properties of polycrystalline titanium nitride (TiN) and aluminum-doped zinc oxide (AZO) by tailoring the film thickness, and their potential for ENZ-enhanced photonic applications. We demonstrate that TiN-AZO bilayers act as Ferrell-Berreman metasurfaces with thickness-tailorable epsilon-near-zero resonances in the AZO films operating in the telecom wavelengths spanning from 1470 to 1750 nm. The bilayer stacks also act as strong light absorbers in the ultraviolet regime employing the radiative ENZ modes and the Fabry-Perot modes in the constituent TiN films. The studied Berreman metasurfaces exhibit optically-induced reflectance modulation of 15% with picosecond response-time. Together with the optical response tailorability of conducting oxides and nitrides, utilizing the field-enhancement near the tunable ENZ regime could enable a wide range of nonlinear optical phenomena, including all-optical switching, time refraction, and high-harmonic generation.

## 1. Introduction

In optical device design, tailoring the optical response of constituent materials is important for achieving the desired functionalities. This makes it important to develop methods of controlling the optical properties of technologically-relevant materials. Titanium nitride is a promising material for nanophotonic applications because of its good plasmonic properties[1], CMOS compatibility[2], tailorability[3,4], versatility of fabrication techniques[5–7], and high laser- and thermal tolerance[8–10]. These attractive features have led to its utilization in high-temperature photovoltaics[11], optical circuitry[12,13], nonlinear optical devices[8,14], and many other practical applications. Similarly, transparent conducting oxides (TCOs) form another class of optical materials for dynamically controlled nanophotonics spanning optical switching[15], electroabsorption modulators[16,17], tunable metasurfaces[18,19], and nonlinear experiments employing epsilon-near-zero (ENZ) physics[20–24]. It is mainly the strong field-enhancement at the ENZ regime that has recently made this class of materials so popular for novel applications that now include ultrafast switching[25–27], tunable broadband light absorption[28–31], enhanced second-harmonic[32] and high harmonic[33] generation.

For these technologically relevant materials, the optical properties can be controlled during growth by varying deposition conditions. For example, the properties of titanium nitride can be tailored by fine-tuning the temperature, gas ratio, and even post-deposition annealing[5,34,35]. The optical properties and the ENZ resonances of conducting oxides can also be altered by varying the growth conditions, namely the gas ratio, the dopant concentrations, or post-deposition annealing[5,17,36,37]. However, a simpler method of tailoring such conducting oxides and nitrides is often overlooked. The conductivity of materials, an intrinsic material property, is strongly affected by the film thickness due to changes in the carrier concentration, crystalline properties, and surface roughness[38–41], especially for thin films. It stands to reason that film thickness should also play a role in the optical property of conducting materials, and in turn, be used to control device characteristics. This work explores the thickness-dependence of the optical properties of polycrystalline titanium nitride (TiN) and aluminum-doped zinc oxide (AZO) and how these materials can be tailored for select passive and dynamic photonic applications, especially those utilizing the ENZ properties of the constituent thin films.

Titanium nitride in its epitaxial form has been utilized in absorbers,[9] waveguiding,[12,13] refractory plasmonics,[8–10] plasmonic nanoparticle lattices,[42] and physics with ultrathin films.[43] However, epitaxial film growth requires lattice-matched substrates, which limits their application in an industry-compatible setting.[44] This difficulty makes it essential to grow and characterize optical-quality films of titanium nitride on non-lattice-matched substrates. We develop low-loss, polycrystalline films of titanium nitride on silicon by reactive magnetron sputtering. Furthermore, we show that the optical properties of the films have a strong dependence on thickness, which adds a way to tailor them. We investigate the cause of the thickness-dependent optical properties through spectroscopic ellipsometry, atomic force microscopy, and transmission electron microscopy, connecting the optical properties to their structural properties. We grow aluminum-doped zinc oxide (AZO) by pulsed laser deposition on the as-grown titanium nitride films. The AZO films show lower losses in the ENZ regimes than films grown by other methods such as sputtering and atomic layer deposition [45,46]. Their optical properties and the ENZ resonances can be tailored during fabrication by varying the thickness.

Building upon the demonstrated adjustability of the TiN and AZO properties, we demonstrate AZO/TiN bilayers where the AZO film is grown on the reflective TiN. Such structures form tailorable, so-called Ferrell-Berreman metasurfaces with the absorption wavelength that can be varied by controlling the AZO thickness. The proposed metasurfaces can operate in a wavelength range from 1470 to 1750 nm, covering the technologically important telecommunication wavelength range. We show that TiN also exhibits both Fabry-Pérot type reflectance dips that vary with the layer thickness and Ferrell-Berreman dips near its ENZ point. As a proof of concept of a dynamic, tunable device, we demonstrate all-optical switching of the metasurface using an interband pump, showing picosecond relaxation times. Understanding the thickness-dependent optical properties of robust, tailorable conducting films will establish a design framework for optimized optical devices[47,48] that could be extended to the important area of the inverse photonic design for globally optimized photonic applications [49,50].

## 2. Tailoring TiN and AZO optical response by varying the thickness

### 2.1. Titanium nitride on silicon

To investigate the effect of thickness on the optical response of polycrystalline TiN films, we grew TiN films of several thicknesses on silicon by DC reactive magnetron sputtering at elevated temperatures. This method was utilized to produce optical grade TiN by other groups [51–53]. Alternative techniques of growing TiN include pulsed laser deposition[42,54] and atomic layer deposition [51]. Supporting Information 1 has details of the growth procedure.

We measured the optical properties of the TiN films with spectroscopic ellipsometry, and fitted them with a Drude-Lorentz model[55], with one free electron term and two Lorentz oscillator terms with the resonance positions at higher energy levels. For the visible wavelengths, the Lorentzian terms govern the response, whereas for longer wavelengths the free electron oscillations become dominant, i.e. the Drude term defines the overall response. The Drude-Lorentz model of the TiN dielectric function is given as follows,

$$\varepsilon = \varepsilon_1 + i\varepsilon_2 = \varepsilon_\infty - \frac{A_0}{(\hbar\omega)^2 + iB_0\hbar\omega} + \sum_{n=0}^{2} \frac{A_n}{E_n^2 - (\hbar\omega)^2 - iB_n\hbar\omega} \quad (1)$$

where $\omega$ is the angular frequency, $\hbar$ is the reduced Planck constant, $\varepsilon_\infty$ is the background permittivity. $A_0$ [eV$^2$] is the square of the plasma frequency and $B_0$ [eV] is the Drude damping coefficient. The Drude term is summed up with two Lorentz terms with $n^{th}$ oscillator strength $A_n$ [eV]$^2$, broadening $B_n$ [eV], and resonance position $E_n$ [eV]. Supporting Information S1 contains the measurement details and the model parameters.

In the telecommunication range, the contribution of free-electrons, represented by the Drude term, strongly governs the optical response. Then, the real and the imaginary parts of the permittivity can be written separately as,

$$\varepsilon_1 = \varepsilon_b - \frac{A_0}{(\hbar\omega)^2 + B_0^2} \quad (2)$$

$$\varepsilon_2 = \frac{A_0 B_0}{\hbar\omega((\hbar\omega)^2 + B_0^2)} \quad (3)$$

where $\varepsilon_b$ denotes the net contribution of the Lorentz oscillators with background permittivity $\varepsilon_\infty$.

The dielectric function of TiN films with different thicknesses are plotted in **Figure 1**.

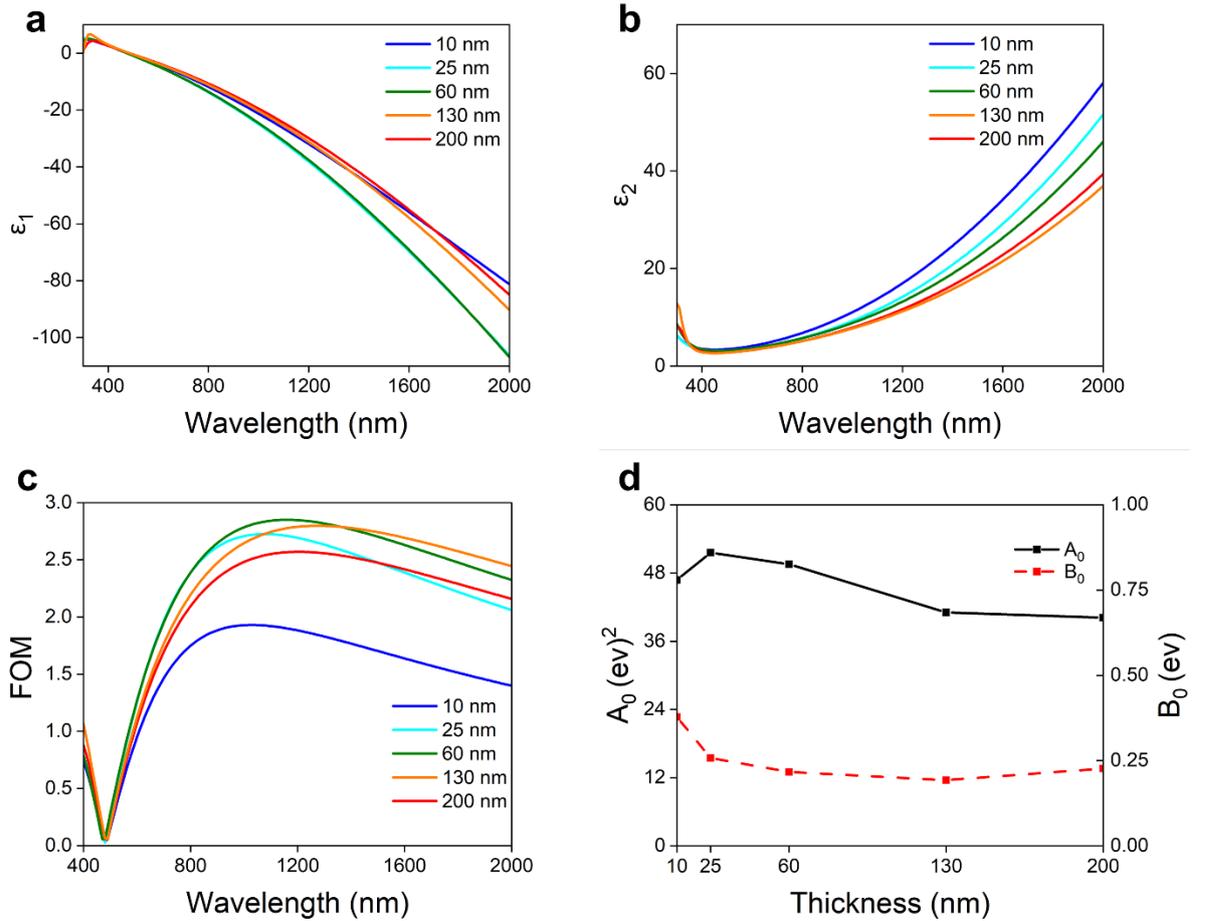

*Figure 1: (a) $\Re(\varepsilon)$ of TiN on Si vs. thickness (b) $\Im(\varepsilon)$ of TiN on Si vs. thickness (c) Plasmonic Figure of Merit vs. thickness (d) Plasma frequency ($A_0$) and Drude damping ($B_0$) vs. the film thickness*

At first, as the thickness increases up to 25 nm, the slope of the real part of the TiN dielectric function becomes steeper, implying increased metallicity (see Fig. 1a). The figure indicates that when the thickness is approaching 60 nm, the change of the real part is saturating. For the imaginary part of the TiN dielectric function ($\varepsilon_2$) this dependence is different; as the film thickness decreases from 130 nm to 10 nm, we observe a monotonic increase that indicates the increasing absorptive losses (Fig. 1b). Figure 1c plots the so-called plasmonic figure of merit (FOM) of the films, defined as the absolute value of the ratio of the real part of the permittivity to the imaginary part.[56,57] The grown TiN films have figure of merits better than the majority of the reported polycrystalline films, and comparable to that of some epitaxial films[1,44,58–62].
To understand the change in the optical response with the thickness, we investigate the crystalline structure of the TiN and its permittivity model. The plasma frequency increases for thicknesses up to 25 nm then starts decreasing (Fig. 1d). An increase in the plasma frequency causes both the real and the imaginary part of the permittivity to increase.
The plasma frequency is related to carrier concentration and the effective mass of the electrons through the equation $\hbar^{-1}\sqrt{A_0} = \omega_p = e\sqrt{\frac{N}{m^*\varepsilon_0}}$. The plasma frequency of polycrystalline films is lower than that of epitaxial titanium nitride films reported in previous studies[13,63]. This is because of the columnar growth of the polycrystalline films seen from TEM (Supporting Information Fig. S1), with many grain boundaries that lower the carrier density[64]. The 10-nm-thick films have the highest damping factors attributed to the increased collisions of electrons with the surface. A similar effect was observed by Shah et al. in a previous study, where the damping factor of TiN thin films increased with decreasing thickness[65]. AFM images (Supporting Information Fig. S2) show that the thicker films have more well-defined and bigger grains. For thicker films, the damping factor generally decreases due to increased grain size, leading to a decreased collision of carriers with the grain boundaries. The films have a surface roughness of less than ~3.5 nm, which is higher than in atomically-flat epitaxial films, yet comparable to polycrystalline films reported previously[42,54]. Overall, the dielectric function of the studied TiN films shows a strong dependence on the thickness, adding a critical degree of control over their optical response.

## 2.2. Al-doped zinc oxide grown on titanium nitride films

We deposited aluminum-doped zinc oxide on optically thick TiN films using pulsed laser deposition (PLD) with a 2% AZO target at a temperature of 120°C. We grew film thicknesses spanning from 27 to 63 nanometers. The 27-43 nm films are grown on 60 nm TiN, and the 57 and 63 nm films are grown on 130 nm TiN. The TEM image of the 43 nm AZO film is given in the Supporting Info S3. For the dielectric functions of AZO, we used a Drude-Lorentz model with a single Lorentz oscillator. Supporting Information S3 contains the model parameters. Figure 2 shows the optical properties of the fabricated AZO films.

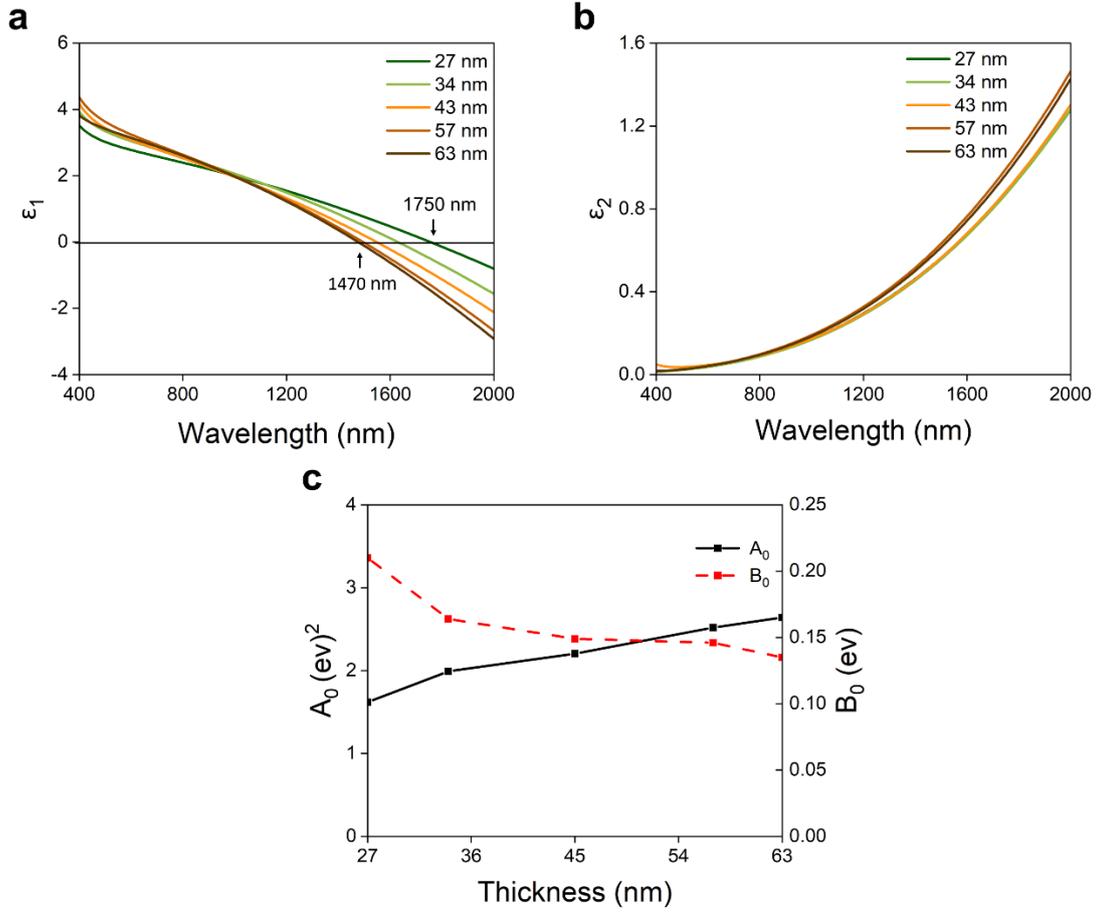

*Figure 2: (a)Real part of permittivity of AZO on TiN vs thickness (b)Imaginary permittivity of AZO vs thickness (c)Plasma frequency ($A_0$) and Drude damping factor ($B_0$) vs thickness*

The ENZ points of the films show a strong blue shift with increasing thicknesses, starting from 1750 nm for the 27-nm-thick films, and reaching the lowest value of 1470 nm for the 63-nm-thick film. The optical losses increase with thickness up to 57 nm then start to saturate for higher thicknesses. With the increasing thickness, the size of the crystalline domains increases, as seen from atomic force microscopy study (Supporting Fig. S3). Thinner films also have more surface defects that can trap electrons and decrease the carrier concentration, and as a result, the plasma frequency increases with the thickness (Fig. 3c)[66–68]. The thicker films have larger grain sizes with a lower surface-to-volume ratio that decreases the scattering channels for electrons, resulting in a decrease in the damping factors with increasing thickness[66–68]. Overall, the thicker films are more metallic and the ENZ frequency blueshifts with increasing thickness.

### 3. Tailoring TiN-AZO bilayer stacks for doubly-resonant Ferrell-Berreman metasurface absorbers

We utilize the thickness-tailorable ENZ points of the conducting films to develop wavelength-selective absorbers. In the epsilon-near-zero regime, materials have diminishingly small electric permittivity that causes various singularities in their optical responses[69]. In this region, strong light-matter interactions induced by slow-light effects and field-intensity enhancement have enabled applications in nonlinearity enhancement [22,23,70], femtosecond optical switching[27,71], time refraction[72–74], optical time reversal[75], and extracting hot electron dynamics in materials [76]. Ultrathin ENZ films demonstrate unique absorption resonances for p-polarized light in their ENZ region, termed the Ferrell-Berreman (FB) modes and the ENZ modes. As the film thickness decreases, the bound surface modes on the upper and the lower interface of the film start interacting with each other, hybridizing into symmetric and antisymmetric modes[77]. Eventually, the symmetric mode forms a flat dispersion band near the ENZ frequency as the thickness is further reduced. This bound resonance is called the ENZ-mode[78,79]. The radiative modes called the Ferrell-Berreman modes are situated in the radiative region occur due to the plasma oscillations in metallic films[80,81], and the excitation of the longitudinal optical phonon in dielectric films[82]. FB resonances have the further advantage of direct excitation from free space without the need for an additional structure enabling low-cost, lithography-free fabrication. ITO[29], CdO[31] and AZO[83] have been recently utilized in applications such as broadband absorption and polarization switching.

A FB metasurface comprises two parts - a robust back-reflector, and an ENZ thin film. The resonance wavelength of the FB metasurface depends on the optical properties of the ENZ film. The TiN-AZO bilayer stacks form a double-resonant FB metasurface, with two dips near the ENZ regions of TiN and AZO for p-polarized light at an angle of incidence of 50 degrees (Fig. 3a). As the AZO ENZ shows a large variance with changing thickness, the FB dip at the telecom frequencies can be tailored by changing the thickness of the films. On the other hand, the TiN ENZ point shows a smaller variance with larger losses, resulting in a FB dip near 480 nm that is not affected by changing the TiN thickness. Near the telecommunication wavelength, the reflectance spectrum for s-polarized light is flat (Supporting Fig. S5). Near the visible wavelengths, s-polarized light shone on TiN shows strong shifts with the thickness variance, indicating the excitation of Fabry-Pérot mode (Fig. 3b). At this wavelength, the AZO layer is almost transparent and the structure can be simplified to a thin TiN layer on a reflective silicon substrate. For p-polarized light, the dip location is proximal to the ENZ wavelength of the TiN film and arises from the excitation of the FB mode in the TiN (Fig. 3c), which is more resistant to changes in thickness. This approach shows how utilizing multilayer stacks and engineering their properties individually can be used to tune the broadband characteristics of the overall structure.

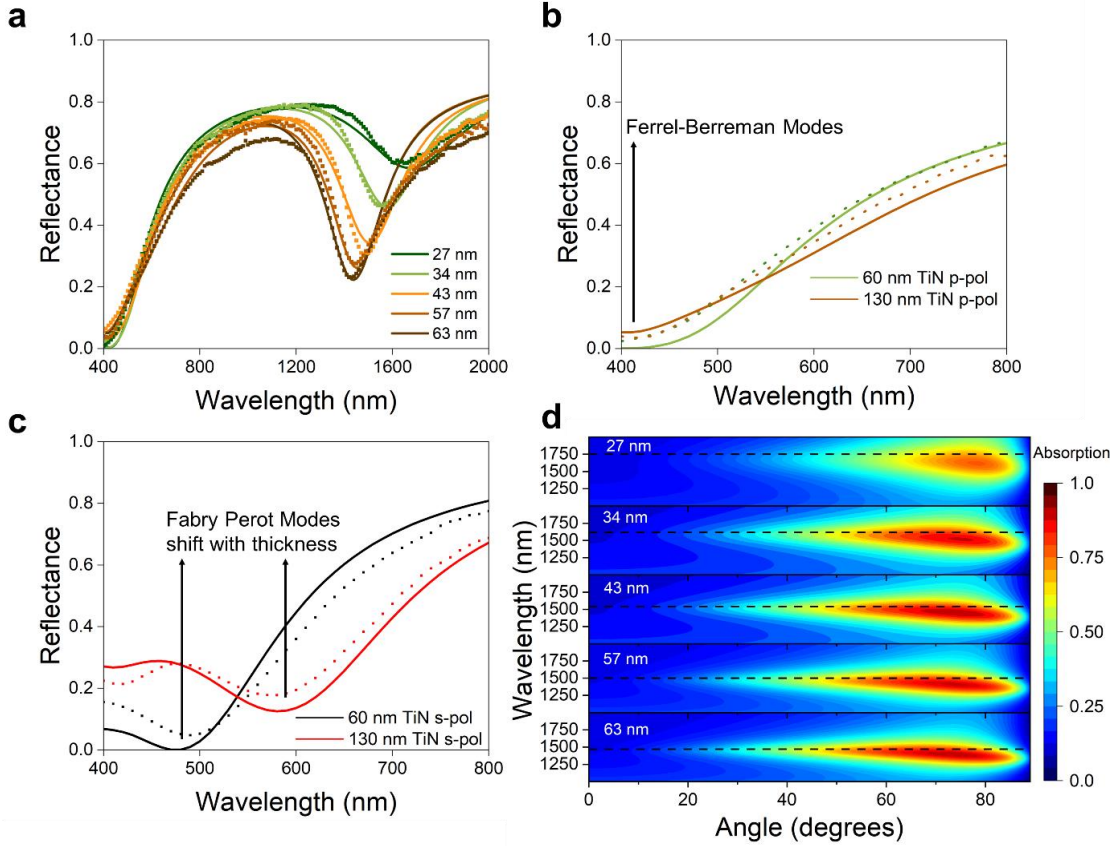

Figure 3: (a) Reflectance Spectrum of the ultra-subwavelength AZO on TiN for p-polarized light. (b) Ferrell-Berreman modes for p-polarized and (c) Fabry-Pérot modes for s-polarized light on the TiN-AZO films with varying TiN thickness (c) Angle and Wavelength Dependent Absorption Spectrum of the AZO films obtained using TMM method

Previously, lithography-free absorbers have been studied for a variety of applications spanning colors[84–86], electrical modulators[87], and polarization switches[26]. To examine the efficacy of the grown films for use in absorbers, we investigated the effect of the angle of incidence on the spectrum theoretically using a Transfer Matrix Method (TMM) model, using the experimentally obtained dielectric properties of the thin films. Figure 3c shows the absorption of the films with respect to the angle of incidence and wavelength. The films show near-perfect absorption around the ENZ region for the longer angle of incidences. The peaks get narrower as the thickness increases because of decreasing losses, desirable for switching applications. Since the absorption is directly related to the intensity enhancement, shown in prior studies by Anopchenko and Gurung et al.[88,89], it would be possible to observe high field intensities inside the AZO film, with potential applications in nonlinear optics spanning time refraction[90], and optical time-reversal[91], and high-harmonic generation[92] using robust ENZ films.

## 4. Exploring TiN-AZO bilayers for dynamic photonics

For a proof-of-concept demonstration of the as-grown films in photonic time-varying applications, we investigated the reflectance modulation of the 63-nm-thick AZO FB absorber with an interband pump – near-infrared (NIR) probe configuration. The pump at 325 nm wavelength is generated by passing a 70 fs, 800 nm pump laser through an OPA. The probe is generated by passing the pump through a sapphire crystal to generate a NIR supercontinuum probe. The pump is at normal incidence, while the probe is at an angle of incidence of 50-degrees (Fig. 4a). Upon excitation by the pump, free carriers generated in the AZO cause the ENZ to blue-shift, resulting in broadband modulation of the reflectance spectrum. As the reflectance spectrum blue shifts due to decreasing refractive index, a positive change is seen to the blue end of the reflectance minimum, and a negative change is seen at the longer wavelengths. A reflectance change of 15% is seen at 1350 nm, at a pump fluence of 1.5 mJ/cm$^2$. The relaxation time of the switching here is on the picosecond scale, with 90% of the signal decaying under 10 ps due to defect-assisted Shockley-Read-Hall mechanisms[27,93], after which the modulation is dominated by slower, thermal effects (Fig. 4b,c)[94]. Active permittivity modulation of AZO with an applied optical pump may be useful for all-optical transistors[95] and other time-varying metasurface applications[90,96,97]. Deeper and faster modulation may be possible employing intraband pumps working at larger angles near the FB resonance[15,98], enabling stronger pump absorption.

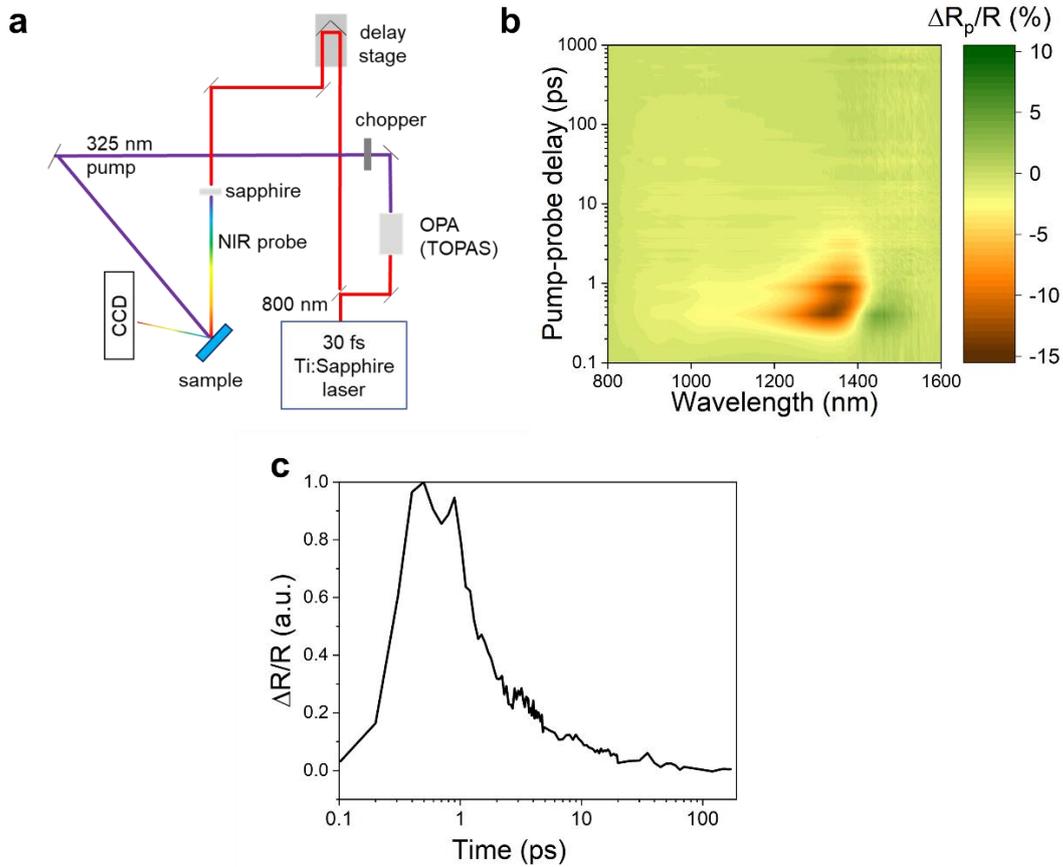

Figure 4: (a) Schematic of the pump-probe setup (b) Color plot showing the reflectance modulation at a pump fluence of 1.5 mJ/cm$^2$, at near infrared wavelengths (horizontal axis) versus time (vertical axis). (c) 90% of the signal decays under 10 ps, after which the relaxation is taken over by slower dynamics of the backreflectors. The plot shows the dynamics at a probe wavelength of 1350 nm.

## 5. Conclusion

In this work, we investigate the thickness-dependent optical response of conducting nitrides and oxides, namely, plasmonic titanium nitride and aluminum-doped zinc oxide as a tailorable, industry-compatible, low-loss platform for passive and dynamic photonics. Specifically, we study the properties of low-loss, optical grade titanium nitride grown on the technologically relevant silicon substrate. The studied films exhibit plasmonic properties that are better than any previously reported work of TiN on silicon, and in some instances, even better than epitaxial titanium nitride grown at lower temperatures. For both TiN and AZO, we demonstrate that by controlling the thickness, the optical properties of the conducting films can be adjusted.

For photonic device demonstrations, we grow aluminum-doped zinc oxide on titanium nitride films with the lowest reported losses in the epsilon near zero regions. Furthermore, we show that the spectral window of the ENZ regime of the studied AZO films can be varied by almost 400 nm by changing the film thickness by ~40 nm, resulting in tailorable Ferrell-Berreman modes spanning the telecommunication wavelength range. Subsequently, we demonstrate strong absorbers utilizing the proposed AZO/TiN bilayers that employ dual Ferrell-Berreman modes, one near the ENZ point of TiN and the other near that of the AZO film. To explore the feasibility of the materials for dynamic photonics, we demonstrate reflectance modulation in the studied AZO/TiN bilayers with pump-probe spectroscopy and show that the absorbance can be tuned at picosecond timescales in the near-infrared range. Since the FB mode only occurs for p-polarized light switching, this can enable active polarization control of the reflected light. The thickness-dependent optical properties of low-loss conducting films can be employed to develop efficient, tailorable photonic devices including lithography-free light absorbers in the near-infrared to the mid-infrared regime, all-optical switches as well as nonlinear structures for high-harmonic generation.

## 6. Acknowledgement

The authors acknowledge support by the U.S. Department of Energy, Office of Basic Energy Sciences, Division of Materials Sciences and Engineering under Award DE-SC0017717 (TCO materials growth and characterization), the Office of Naval Research under Award N00014-20-1-2199 (TCO optical characterization) and the Air Force Office of Scientific Research under Award FA9550-20-1-0124 (transition metal nitrides studies). Use of the Center for Nanoscale Materials, an Office of Science user facility, was supported by the U.S. Department of Energy, Office of Science, Office of Basic Energy Sciences, under Contract No. DE-AC02-06CH11357.

**Supporting Information for *"Tailoring the Thickness-Dependent Optical Properties of Conducting Nitrides and Oxides for Epsilon-Near-Zero-Enhanced Photonic Applications"***

## S1. Growth of TiN and measurement of the optical properties

TiN growth by reactive DC magnetron sputtering (PVD Products). A 99.995% pure Ti target of 2-inch diameter was used for the process. To ensure uniformity, the distance from the target to the source is kept at 20 cm. The chamber is pumped down to $10^{-8}$ T to prevent oxygen contamination, and is backfilled with Ar to a pressure of 5 mTorr. The Ti is sputtered for 2 minutes to clean the top surface with a power of 200W, and then an Ar:$N_2$ mixture of 1:18 is used for the sputtering process. The substrate is heated to a temperature of 800°C and rotated at 5 rpm. We note that it is possible to grow epitaxial titanium nitride on silicon, but complicated cleaning procedures are required to remove the native oxide layer[1]. For this study, we wanted to focus on the evolution of the optical properties of polycrystalline titanium nitride with thickness.

The films are characterized by variable angle spectroscopic ellipsometry (VASE) from 300 to 2000 nm wavelength employing JA Woollam W-VASE ellipsometer, at angles of incidences of 50 and 70 degrees. The thicknesses of the TiN films are set as constants in the fit. For the silicon substrate, a thickness of 0.5 mm is used, and we used the model described by Herzinger et al[2].

The dielectric function of transition metal nitrides such as TiN can be modeled with a Drude-Lorentz model.[3] The interband transitions are typically described by the Lorentz Oscillator model and the free-electron response by the Drude model as follows,

$$\varepsilon = \varepsilon_1 + i\varepsilon_2 = \varepsilon_\infty - \frac{A_0}{(\hbar\omega)^2 + iB_0\hbar\omega} + \sum_{n=0}^{2} \frac{A_n}{E_n^2 - (\hbar\omega)^2 - iB_n\hbar\omega} \qquad (S1)$$

Table S1 comprises all the Drude-Lorentz parameters of the films. Incorporating the surface roughness into the top surface of the TiN layer employing effective medium approximation generally decreased the MSE by a negligible amount even for the films with 3.4 nm roughness. As a result, we excluded the roughness from the fitting.

*Table S1: The Drude-Lorentz Model Parameters of TiN Films on Si vs. TiN Thickness*

| Thickness | 10 nm | 25 nm | 60 nm | 130 nm | 200 nm |
|---|---|---|---|---|---|
| $A_0$ | 46.782 | 51.595 | 49.550 | 41.055 | 40.152 |
| $B_0$ | 0.378 | 0.258 | 0.217 | 0.193 | 0.227 |
| $A_1$ | 98.232 | 98.232 | 52.954 | 31.801 | 28.517 |
| $B_1$ | 2.157 | 2.157 | 1.364 | 0.661 | 1.003 |
| $E_1$ | 5.119 | 5.119 | 4.466 | 4.117 | 4.117 |
| $A_2$ | 407.280 | 407.280 | 310.930 | 357.600 | 391.660 |
| $B_2$ | 149.950 | 149.950 | 79.364 | 80.546 | 99.451 |
| $E_2$ | 5.982 | 5.982 | 4.044 | 3.945 | 4.748 |
| $\varepsilon_\infty$ | 2.26 | 2.84 | 3.39 | 3.20 | 3.41 |
| MSE | 5.8 | 2.8 | 4.1 | 8.9 | 8.7 |

## S2. The crystalline structure of the as-grown films measured by Atomic Force Microscopy and TEM

The crystalline structures of the film also change with the thickness which can be used to explain the change in the optical parameters. Figure S1 shows the AFM image of the TiN films, which suggests the grain size and roughness increase with increasing thickness. Grain boundaries and different crystal orientations are visible from the TEM image showing the polycrystalline growth of the TiN (Fig S2).

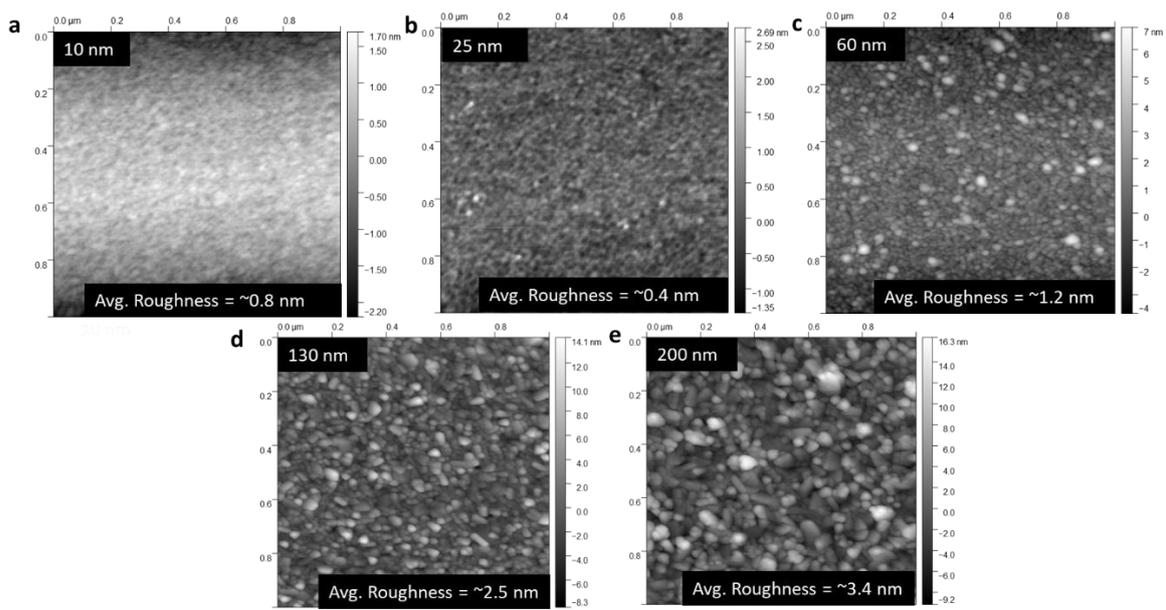

*Figure S1: AFM image of TiN films with different thicknesses of (a) 10 nm, (b) 25 nm, (c) 60 nm, (d) 130 nm and (e) 200 nm. The roughnesses vary from 0.4 to 3.4 nm.*

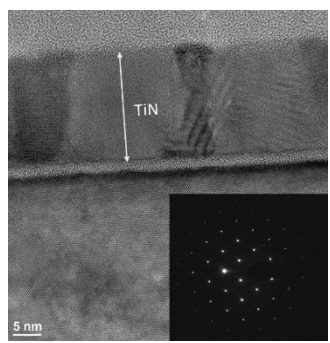

*Figure S2: TEM image of the 60 nm TiN film, showing the polycrystallinity. The inset shows the diffraction pattern of the film.*

## S3. Growth of AZO and extraction of the optical properties

AZO is grown on 60 nm TiN films by pulsed laser deposition (PVD-Products). The substrate is heated at a temperature of 40 C. A 2% AZO target is irradiated with a 5 Hz 266 nm Excimer laser with a fluence of 1 J/cm$^2$/pulse. To verify the thicknesses, we first grow a ~400 nm thick film of AZO on silicon and measure the thickness through cross-sectional SEM. The growth rate is estimated from the thickness divided by time. The deposition rate is 6 nm/min. Different thicknesses of films were then grown by adjusting the deposition time.

Spectroscopic ellipsometry is done on the AZO samples at angles of 50 and 70 degrees and fitted with a Drude-Lorentz model. The thickness for AZO was added as a fitting parameter in the JA W_VASE software, and the extracted thickness was verified by tunneling electron microscopy. Fig. S4 has the crystalline profile of the 43 nm sample to serve as an example. Table S2 has the table of the parameters. The crystallinity of the AZO is investigated using AFM (Figure S3) and TEM (Figure S4) measurements.

*Table 2: The Drude-Lorentz Parameters of AZO Films on TiN vs. AZO thickness*

| Thickness | 27 nm | 34 nm | 45 nm | 57 nm | 63 nm |
| --- | --- | --- | --- | --- | --- |
| $A_0$ | 1.619 | 1.990 | 2.205 | 2.519 | 2.64 |
| $B_0$ | 0.210 | 0.164 | 0.149 | 0.146 | 0.135 |
| $A_1$ | 9.794 | 16.502 | 18.700 | 21.544 | 7.412 |
| $B_1$ | 0.005 | 0.006 | 0.050 | 0.006 | 0.019 |
| $E_1$ | 4.143 | 4.368 | 4.282 | 4.386 | 4.258 |
| $\varepsilon_\infty$ | 2.392 | 2.392 | 2.264 | 2.392 | 3.224 |
| MSE | 16.64 | 22.18 | 31.84 | 24.89 | 22.29 |

Figure S2 shows the AFM image of the AZO films. The grain size increases with thickness, and thicker films have better-defined grains.

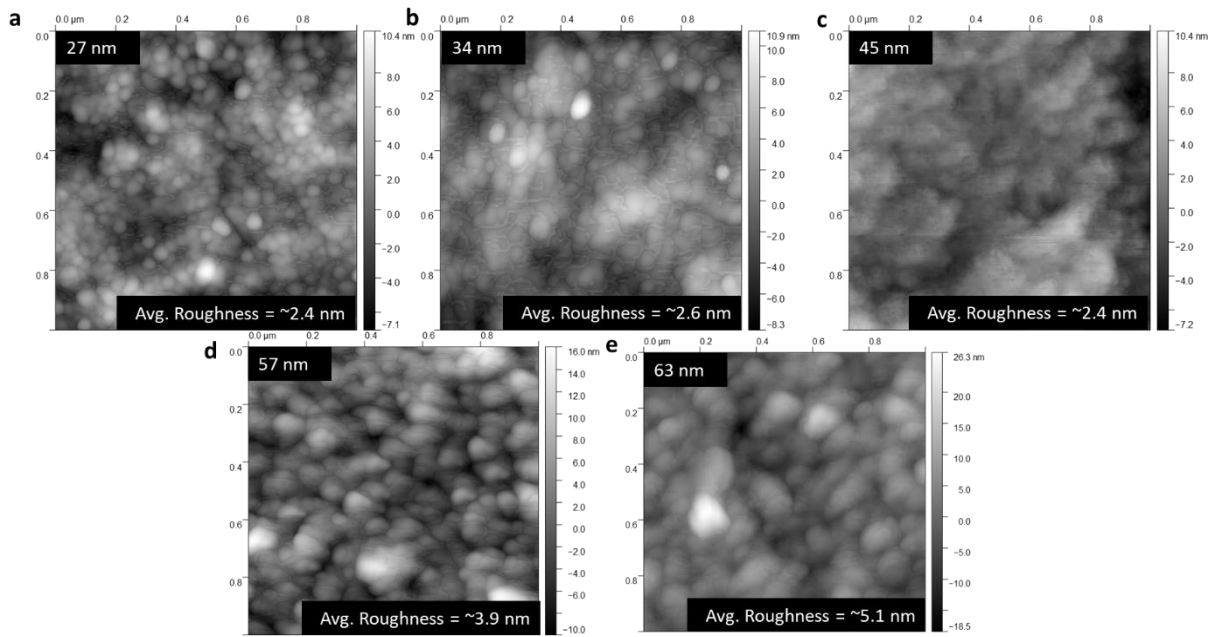

*Figure S3: AFM image of AZO films of thickness (a) 27nm, (b) 34 nm, (c) 45 nm, (d) 57 nm and (e) 63 nm, grown on TiN, showing the increasing grain sizes with thickness*

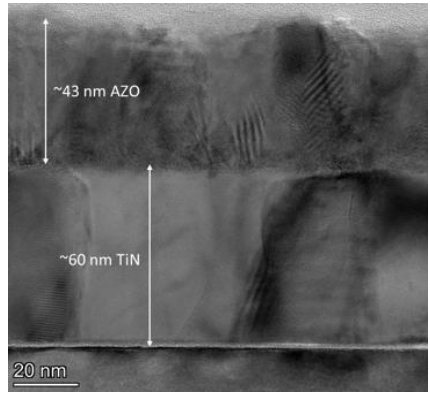

*Figure S4: TEM image of the 43 nm AZO film on 60 nm TiN. Both TiN and AZO films are polycrystalline*

## S4. Steady-state response of absorber for s-polarized light shows no dips near the ENZ of AZO

We performed reflectance measurements on the films at an angle of 50° for the AZO films on optically thick TiN. The spectrum of the s polarized light is primarily flat near the telecommunication wavelengths, with dips near the visible wavelength regime because of the Fabry Perot resonances. Near the telecommunication wavelength, the graph is flat for s-polarization, a feature of the Ferrell Berreman mode (Figure S5).

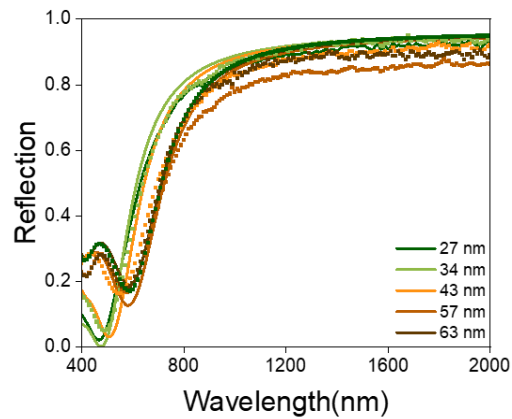

*Figure S5: Reflectance Spectrum of the AZO films for s polarized light. The solid lines represent the fitted graphs from the Drude Lorentz model, and the dots represent the experimental results. There are no dips near the ENZ of the AZO films, a distinguishing feature of the Berreman mode*